\documentclass{elsart5p}

\pdfoutput=1
\usepackage{amsmath}
\usepackage{amsfonts}
\usepackage{amssymb}
\usepackage{graphicx}

\renewcommand{\Re}{\mathrm{Re}}

\begin{document}

\begin{frontmatter}

\title{Desynchronization of large-scale neural networks by stabilizing unknown unstable incoherent equilibrium states}

\author{Tatjana Pyragien{\.e} and Kestutis Pyragas}
\address{Center for Physical Sciences and Technology, Sauletekio al. 3, LT-10257 Vilnius, Lithuania}

\begin{abstract}
In large-scale neural networks, coherent limit cycle oscillations usually coexist with unstable incoherent equilibrium states, which are not observed experimentally. We implement a first-order dynamic controller to stabilize unknown equilibrium states and suppress coherent oscillations. The stabilization of incoherent equilibria associated with unstable focus and saddle is considered. The algorithm is demonstrated for networks composed of quadratic integrate-and-fire (QIF) neurons and Hindmarsh-Rose neurons.  The microscopic equations of an infinitely large QIF neural network can be reduced to an exact low-dimensional system of mean-field equations, which makes it possible to study the control problem analytically.
\end{abstract}


\end{frontmatter}

\section{Introduction}\label{Intro}

Synchronization studies in large populations of coupled oscillatory or excitable elements are relevant in fields ranging from physics to neuroscience \cite{winf01,Kuramoto2003,pikov01,Boccaletti2018}.  The role of synchronization in neural systems can be twofold. In a healthy state, it is responsible for learning and cognition ~\cite{Singer1999,Fell2011}, however, excessive synchronization can cause a variety of neurological conditions such as Parkinson's disease~\cite{Hammond2007}, epilepsy~\cite{Jiruska2013,Gerster2020}, tinnitus~\cite{Tass2012tin}, and others. High-frequency (HF) deep brain stimulation (DBS) is a standard procedure for the treatment of neurological disorders~\cite{Benabid1991,kring07}. The mechanisms of DBS are not yet well understood~\cite{Lozano2004,Lozano2019}. Simple models show that the HF DBS effect can be explained either as the result of stabilizing the resting state of individual neurons~\cite{Pyragas2013} or as suppressing synchronized oscillations without forcing individual neurons into silence~\cite{Pyragas2021}. HF DBS may cause side effects and its therapeutic effect may decrease over time, so there is a significant clinical need for less invasive and more effective stimulation methods~\cite{Krauss2021}. In open loop control systems such as HF DBS, adverse effects on neural tissue can be reduced by optimizing the waveform of the stimulus signal~\cite{Pyragas2018,Pyragas2020}. 

However, a number of theoretical works show that the desynchronization of coherent oscillations is especially effective with the help of closed-loop (feedback) control algorithms. Various control strategies based on linear \cite{rosenblum04a,rosenblum04b,Hauptmann05a,Hauptmann05b,batista10} and nonlinear \cite{Popovych05,Popovych06,Popovych10} time-delayed feedback, linear feedback bandpass filters \cite{Tukhlina2007,ming09,Montaseri13}, proportional-integro-differential feedback with a separate stimulation-registration setup \cite{Pyragas2007},  act-and-wait time-delayed feedback \cite{Ratas2014,Ratas2016ND}  and others \cite{Rosin11,Berenyi12,Louzada12}  were considered.

Recent advances in the theory of nonlinear dynamical systems have provided the neuroscience community with simple, low-dimensional models of neural networks referred to as next-generation neural mass models~\cite{Coombes2023}. Such models are useful objects for developing, testing, and understanding various synchronization control algorithms. Here we show that these models can naturally explain the desynchronization mechanism of our feedback control algorithm in terms of stabilizing unknown unstable incoherent states. The next-generation models are derived directly from the microscopic dynamics of individual neurons and are accurate in the thermodynamic limit of infinite network size. These models represent a closed system of mean-field equations for biophysically relevant parameters such as mean membrane potential and firing rate.  Low-dimensional dynamics in a large population of coupled oscillatory elements was first discovered by Ott and Antonsen~\cite{Ott2008} in the Kuramoto model~\cite{Kuramoto2003}. Later, this discovery was successfully applied to derive a low-dimensional system of mean-field equations for a certain class of networks consisting of all-to-all pulse-coupled QIF neurons~\cite{Montbrio2015}, which are canonical models of class I  neurons~\cite{ermentrout10}. 

In recent years, next-generation models have been obtained for a large number of different modifications of QIF neural networks 
\cite{Pazo2016,Ratas2016,Ratas2018,Ratas2019,Montbrio2020,Segneri2020,GoldobinPRL2021,Goldobin2021,PyragasV2021,PyragasV2022,PyragasV2023}. 
These models make it possible to carry out their detailed bifurcation analysis and reveal synchronization mechanisms. It has been shown that synchronized limit cycle oscillations can arise from various bifurcations, such as the Hopf bifurcation in Refs.~\cite{Pyragas2021,Ratas2016} or the homoclinic bifurcation in Ref.~\cite{Ratas2016}. However, stable limit cycles are always accompanied by unstable fixed points, which correspond to unstable incoherent equilibrium states of the network. These unstable states are not observed experimentally. Here, we show that a priori unknown unstable incoherent states can be stabilized using the control algorithm proposed in Refs.~\cite{Pyragas2002,Pyragas2004}. Initially, this algorithm was developed and tested to stabilize unknown unstable equilibrium states of low-dimensional dynamical systems, and recently it has been implemented to stabilize unstable pedestrian flows in the collective behavior of large crowds of people~\cite{Just2022}. Here, we implement this algorithm to stabilize unstable incoherent states in large-scale neural networks consisting of QIF and Hinmarsh-Rose \cite{Hindmarsh1984} neurons. We demonstrate effective control of two types of equilibrium states associated with an unstable focus and a saddle point. As far as we know, the control of saddle equilibrium states in neural networks has not been considered in the literature.

The paper is organized as follows.  Section~\ref{sec:control algorithm} describes the control algorithm. In Sec.~\ref{sec:QIFone}, we apply this algorithm to a  population of synaptically coupled excitatory QIF neurons. Here we stabilize incoherent states associated with an unstable focus and a saddle fixed point. The latter is stabilized by an unstable controller. Section~\ref{sec:QIFtwo} is devoted to the control of two interacting populations of excitatory and inhibitory QIF neurons. In Sec.~\ref{sec:HindMarshRose}, we apply our algorithm to a population of chaotically spiking Hindmarsh-Rose neurons, whose microscopic model equations cannot be reduced to a low-dimensional system. 
The conclusions are presented in Sec.~\ref{sec:Conclusions}.

\section{Control algorithm}\label{sec:control algorithm}

We consider a large network of coupled neurons generating collective coherent oscillations. We assume that, along with the synchronous mode of coherent oscillations, the network has an unstable equilibrium state characterized by incoherent oscillations of individual neurons. Our goal is to stabilize the incoherent state and transition the network from synchronous to incoherent mode. To achieve this goal, we turn to the algorithm for stabilizing unknown unstable equilibrium points of low-dimensional dynamical systems, developed in Refs.~\cite{Pyragas2002,Pyragas2004}. The algorithm uses a simple first-order dynamic controller based on a low-pass filter (LPF). The block diagram of this algorithm, adapted for neural networks, is shown in Fig. 1. 
\begin{figure}
\centering
\includegraphics{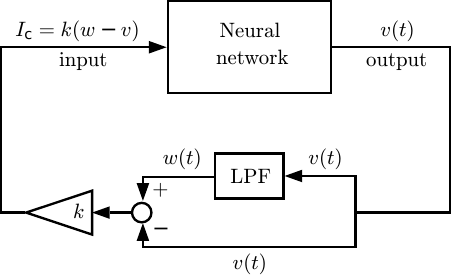}
\caption{Block diagram of stabilization of unknown incoherent states in neural networks. The mean membrane potential $v(t)$ represents the output of the network.
The network is stimulated by the input current $I_c(t)$. In a feedback loop, LPF stands for low-pass filter.}
\end{figure}

We assume that the mean membrane potential $v(t)$ of the entire or some part of the neural population can be measured at the output of the network. In addition, we assume that all or part of the population of neurons can be stimulated by the input current $I_c(t)$. In general, the measured and stimulated subpopulations may differ. The input and output of the network are connected by a feedback loop described by the following equations:
\begin{subequations}\label{controllaw}
\begin{eqnarray}
    \dot{w}&=&\omega_c(v-w),\label{controllawa} \\
    I_c&=&k(w-v),\label{controllawb} 
\end{eqnarray} 
\end{subequations}
where $w$ is a dynamic variable of the controller (LPF). The control algorithm has two adjustable parameters: the cutoff frequency $\omega_c$ of the LPF and the control gain $k$. Let us denote the average membrane potential of the free network in a state of unstable equilibrium as $v=v^*$, which in the thermodynamic limit should be a constant, $v^*=const$. We assume that this value is a priori unknown. The control algorithm is designed in such a way that the equilibrium value of $v^*$ remains unchanged in the stationary state of the closed loop system. Indeed, at $\dot{w}=0$ the control variable coincides with the mean membrane potential $w=w^*=v^ *$, and the feedback perturbation vanishes, $I_c=0$. However, feedback perturbation affects the stability of the incoherent state. The examples below show that this state can be stabilized by adjusting control parameters $\omega_c$ and $k$ accordingly.

This algorithm has a number of advantages. Firstly, it is weakly invasive.  Below, we will show that the feedback perturbation $I_c$ decreases according to a power law with increasing network size and vanishes as the network size tends to infinity. Secondly, this algorithm does not require knowledge of the mean membrane potential $v^*$ of an unstable equilibrium state and, thirdly, the algorithm provides tracking of the equilibrium state in the case of slowly varying system parameters~\cite{Pyragas2004}.

Note that the control algorithm with an ordinary LPF ($\omega_c>0$) has a limitation. It works well for unstable equilibrium points like focuses but doesn't work for saddles. More precisely, Ref.~\cite{Pyragas2002} gives a theorem that a stable controller cannot stabilize unstable equilibrium points with an odd number of real positive eigenvalues. This limitation can be avoided by using an unstable controller in the same way as it is done in the delayed feedback control algorithm~\cite{Pyragas1992} when stabilizing a certain type of unstable periodic orbits~\cite{Pyragas2001}. Here, to stabilize an unstable incoherent state of the saddle type, we will use an unstable LPF with the parameter $\omega_c<0$. An unstable LPF can be implemented using an RC circuit with a negative resistor.

In the following sections, we will demonstrate the performance of this algorithm for three examples of neural networks. The first two examples deal with large populations of synaptically coupled QIF neurons. In the limit of infinite size, microscopic models of these networks can be reduced to exact low-dimensional systems of mean-field equations. In the first example, one population of excitatory neurons is considered, and in the second example, two interacting populations of excitatory and inhibitory neurons are analyzed. The third example is devoted to electrically coupled chaotic Hindmarsh-Rose neurons.

\section{Controlling a population of synaptically coupled  excitatory QIF neurons} \label{sec:QIFone}

First, we apply the algorithm described above to a heterogeneous population of QIF excitatory neurons interacting via finite-width synaptic pulses~\cite{Ratas2016}. The microscopic state of the population is defined by the set of $N$ neurons' membrane potentials  $\{V_j \}_{j=1,\ldots,N}$. They satisfy the following set of equations~\cite{ermentrout10}: 
\begin{eqnarray}
\dot{V}_{j} &=&  V_{j}^{2}+\eta_j+Js(t)+I_c(t),\;\;\nonumber\\
& & \;\; \text{if} \; V_j \ge V_p \; \text{then} \; V_j \leftarrow V_{r}. \label{model}
\end{eqnarray}
Here, $\eta_j$ is a heterogeneous excitability parameter that specifies the behavior of individual neurons and the term $Js(t)$ stands for the synaptic coupling, where $J$ is the synaptic weight and $s(t)$ is the normalized mean synaptic current emitted by spiking neurons. The term $I_c(t)$ describes an external current, which we interpret as a control variable.  In this model, the membrane time constant of QIF neurons is assumed to be unity. This means that time here is measured in units of the membrane time constant.

For $J=0$ and $I_c=0$, the neurons with  the parameter $\eta_j<0$ are at rest, and the neurons with the  parameter $\eta_j>0$ generate spikes. When the potential $V_j$ reaches the threshold value $V_p$, it is instantly reset to the value $V_r$. We choose thresholds in the form $V_{p}= -V_{r} = \infty$, which allows us to transform QIF neurons into theta neurons and obtain an accurate system of reduced mean-field equations ~\cite{Montbrio2015}.  We consider the case when the heterogeneous parameter $\eta$ is distributed according to the Lorentzian density function
\begin{equation}\label{lor}
    g(\eta)=\frac{1}{\pi}\frac{\Delta}{(\eta-\bar{\eta})^{2}+\Delta^{2}},
\end{equation}
where $\Delta$ is the half-widths  and  $\bar{\eta}$ is the center of the distribution. For the Lorentzian heterogeneity, the reduction of microscopic equations is the most efficient. Note that other distributions of the heterogeneous parameter have been considered in recent publications.~\cite{PyragasV2021,PyragasV2022}.

Here we use the model of global coupling in which neurons emit synaptic pulses of finite width with the mean synaptic current defined as~\cite{Ratas2016}
\begin{equation}
	s(t)=\frac{V_{th}}{N}\sum_{i=1}^{N}H(V_i(t)-V_{th}), \label{ssyn}
\end{equation}
where $H(\cdot)$ is the Heaviside step function and $V_{th}$ is a threshold potential that determines the height and width of synaptic pulses. 

In the limit $N\to \infty$, the above microscopic model reduces to an exact system of two ordinary differential equations (ODEs)~\cite{Ratas2016}
\begin{subequations}
\label{eqrv}
\begin{eqnarray}
\dot{r} & = & \Delta/\pi+ 2rv,\label{eqrv1}\\
\dot{v} & = & \bar{\eta} +v^2-\pi^2 r^2+Js(t)+I_c(t)\label{eqrv2}
\end{eqnarray}
\end{subequations}
for two biophysically relevant parameters, the mean spiking rate  $r(t)$ and the mean membrane potential $v(t)$. In the infinite size limit, the mean synaptic current~(\ref{ssyn}) is expressed in terms of the parameters $r(t)$ and $v(t)$ as~\cite{Ratas2016}
\begin{equation}
s(t)=\frac{V_{th}}{\pi} \left[\frac{\pi}{2}- \arctan \left(  \frac{V_{th}-v(t)}{ \pi r(t) }  \right)\right]. \label{ssynmacro}
\end{equation}
This expression closes the system of mean-field Eqs.~(\ref{eqrv}). The bifurcation analysis of these equations without control $I_c(t)=0$ was carried out in Ref.~\cite{Ratas2016}. This analysis showed that synchronous limit cycle oscillations can occur through two types of bifurcations: the Hopf bifurcation and the homoclinic bifurcation. In the first case, the system~(\ref{eqrv}) has a stable focus before the bifurcation. On a microscopic level, this corresponds to a stable equilibrium state of the network with incoherent dynamics of individual neurons. After the bifurcation, the incoherent equilibrium state becomes an unstable focus, and neurons exhibit coherent limit cycle oscillations. Our goal here is to bring back the incoherent dynamics by stabilizing the unstable equilibrium state. In the case of a homoclinic bifurcation, the limit cycle touches the saddle point and becomes a homoclinic orbit. Near this bifurcation, we will suppress coherent oscillations by using an unstable controller to stabilize the incoherent state of the saddle equilibrium.

We begin the application of our control algorithm from the case of limit cycle oscillations arising from the Hopf bifurcation. We use typical system parameters corresponding to this mode~\cite{Ratas2016}: $\Delta=1$, $V_{th}=50$, $\bar{\eta}=2$, and $J=20$. For these parameters, the only attractor in the two-dimensional phase space ($r,v$) of the free ($I_c=0$) system~(\ref{eqrv}) is the limit cycle. Inside this cycle there is an unstable focus with the coordinates 
$(r^*, v^*) \approx (2.1081, -0.0755)$ and two complex-conjugate eigenvalues $\lambda_{1,2} \approx 0.2621 \pm 9.6190$. Let us now estimate how the local properties of this fixed point change in the presence of a control defined by the Eq.~(\ref{controllaw}). Due to the additional variable $w$, the phase space of the closed loop system is expanded to three dimensions: $(r,v,w)$. The coordinates of the fixed point in the three-dimensional phase space are $(r^*,v^*,v^*)$, i.e. its projection onto the original two-dimensional phase space remains unchanged. However, the stability properties of this fixed point now depend on the controller parameters $\omega_c$ and $k$ and are determined by the eigenvalue problem
\begin{equation} \label{Eigen}
\det (A-\lambda  I)=0
\end{equation} 
of the linearized system of Eqs.~(\ref{eqrv}) and 
(\ref{controllaw}). Here 
\begin{equation} \label{Jacob}
A=
\begin{pmatrix}
a_{11} & a_{12} & 0 \\ 
a_{21} & a_{22}-k & k \\ 
0 & \omega_c & -\omega_c
\end{pmatrix}
\end{equation} 
is the Jacobian matrix of this system, $a_{ij}$  are the coefficients of the Jacobian matrix of of the system~(\ref{eqrv}) without control evaluated at the fixed point $(r^*,v^*)$. Specifically, $ a_{11}=2v^*$, $a_{12}=2r^*$, $a_{21}=-2\pi^2r^*+JV_{th}(\pi r^*)^{-2}(V_{th}-v^*)c^{-1}$ and $a_{22}=2v^*+JV_{th}\pi^{-2}(c r^*)^{-1}$, where $c=1+\left[(V_{th}-v^*)/(\pi r^*)\right]^2$. Finally, $I$ is the identity matrix, and $\lambda$ is the eigenvalue. 

\begin{figure}
\centering\includegraphics{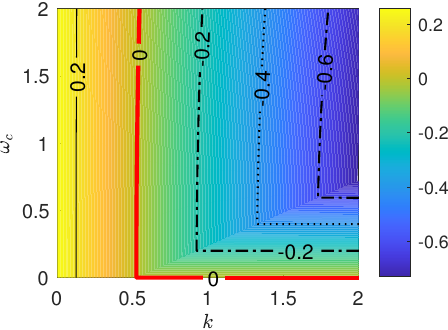}
\caption{The performance of the control algorithm depending on the control parameters $\omega_c$ and $k$. The results for an unstable focus in a population of synaptically coupled QIF neurons are presented. The contour lines and colors indicate the maximum real part of the eigenvalues $\max[\Re(\lambda)]$ obtained from the Eq.~(\ref{Eigen}). The thick red contour line corresponds to $\max[\Re(\lambda)]=0$. It separates stable and unstable regions. The originally unstable focus is stabilized in the region $\max[\Re(\lambda)]<0$. Network parameters: $\Delta=1$, $V_{th}=50$, $\bar{\eta}=2$, and $J=20$.}
\label{Focuslamkw}
\end{figure}

For a given fixed point, the dependence of the solutions of the Eq.~(\ref{Eigen}) on the parameters $\omega_c$ and $k$ is shown in Fig.~\ref{Focuslamkw}. 
The colors encode the values of $\max[\Re(\lambda)]$. The thick red contour line corresponds to $\max[\Re(\lambda)]=0$. It separates regions of a stable and unstable fixed point. We see that the control algorithm is robust to the choice of control parameters $\omega_c$ and $k$. The algorithm provides stabilization of the unstable focus for any $\omega_c>0$ and $k \gtrapprox 0.55$. 
 
\begin{figure}
\centering\includegraphics{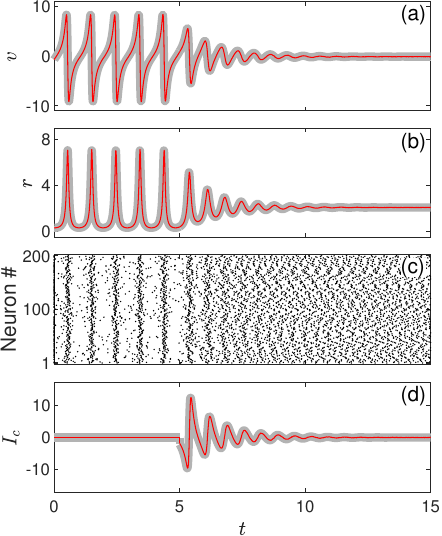}
\caption{Suppression of coherent oscillations by stabilization of an unstable focus in a population of synaptically coupled QIF neurons. For $t<5$, there is no control and the network generates collective coherent oscillations. For $t>5$, the control is turned on and the system goes into a previously unstable incoherent state. The dynamics of (a) mean membrane potential, (b) mean spiking rate, and (d) control perturbation derived from the mean-field Eqs.~(\ref{eqrv}) are shown as thick gray curves. The thin red curves show the same results derived from the microscopic model~(\ref{thetaj}). (c) Raster plot of $200$ randomly selected neurons. The spike moments for each neuron are shown by dots. The neuron numbers are shown on the vertical axis.  The parameters of the network are the same as in Fig.~\ref{Focuslamkw}. Controller parameters: $\omega_c=1$ and $k=2$. The microscopic model was simulated using $N=10^4$ neurons. 
} 
\label{fig:Focusmicromacrodyn}
\end{figure}

Figure~\ref{fig:Focusmicromacrodyn} shows the performance of the control algorithm for fixed values of $\omega_c=1$ and $k=2$. 
The thick gray curves show the dynamics of the free and controlled neuronal population obtained from the mean-field equations ~(\ref{eqrv}). During the time $t<5$ the control is switched off and the system is in the mode of limit cycle oscillations. The mean membrane potential [Fig.~\ref{fig:Focusmicromacrodyn}(a)] and the mean spiking rate [Fig.~\ref{fig:Focusmicromacrodyn}(b)] show periodic oscillations. At $t>5$ the control is activated and the oscillations are damped. The system approaches a stabilized equilibrium state. The control perturbation [Fig.~\ref{fig:Focusmicromacrodyn}(d)] experiences transient damped oscillations and vanishes asymptotically.

As a next step, we tested the performance of our algorithm for networks of finite size, described by the microscopic Eqs.~(\ref{model}). Unlike the low-dimensional mean-field Eqs.~(\ref{eqrv}), the microscopic model is defined by a huge number of differential equations. The typical population sizes we model here are $N \sim 10^4$ neurons. There is of course no {\it a priori} guarantee whether the control algorithm will work for such high-dimensional systems. Numerical simulation is more convenient after changing variables
\begin{equation}
V_j = \tan(\theta_j/2),\label{eqtransftet}
\end{equation}
which transforms QIF neurons into theta neurons. The advantage of theta neurons is that they avoid the discontinuty  problem. When the membrane potential $V_j$ of the QIF neuron rises to $+\infty$ and falls to $-\infty$, the theta neuron simply crosses  the phase  $\theta_j=\pi$. For theta neurons, the Eqs~(\ref{model}) are transformed to
\begin{eqnarray}
\dot{\theta}_{j} &=& 1-\cos \left(\theta_{j}\right)\;\;\nonumber\\
&+& \left[1+\cos \left(\theta_{j}\right)\right]\left[\eta_{j}+J s(t)+I_c(t) \right]. \label{thetaj}
\end{eqnarray}
We integrated these equations by the Euler method using a time step of $d t = 10^{-4}$. We have generated the values of the Lorentzian distributed (\ref{lor}) heterogeneous parameter deterministically using $\eta_j=\bar{\eta}+\Delta \tan(\pi/2)(2j-N-1)/(N+1)]$ for   $j=1,\ldots, N$. For more details on modelling the Eqs.~(\ref{thetaj}), see Ref.~\cite{Ratas2016}. From the Eqs. ~(\ref{thetaj}), we estimated the Kuramoto order parameter~\cite{Kuramoto2003}
\begin{equation}
\label{eqZ}
Z=\frac{1}{N}\sum\limits_{j=1}^{N}\exp(i \theta_j)
\end{equation} 
and used its relation with the spiking rate $r$ and the mean membrane potential $v$~\cite{Montbrio2015}: 
\begin{equation}
\label{eqW}
r=\frac{1}{\pi}\operatorname{Re}\left(\frac{1-Z^*}{1+Z^*}\right), \quad v=\operatorname{Im}\left(\frac{1-Z^*}{1+Z^*}\right),
\end{equation} 
where $Z^*$ denots complex conjugate of $Z$. 

Results derived from the microscopic model~(\ref{thetaj}) for $N=10^4$ neurons are presented in Fig.~\ref{fig:Focusmicromacrodyn} by thin red curves. They are in good agreement with the results obtained from the reduced mean-field Eqs.~(\ref{eqrv}). Thus, the control algorithm works well for a large population of $N=10^4$ neurons, and the mean-field theory correctly predicts the dynamics of the population in the presence of control. To demonstrate network dynamics at the microscopic level, Fig.~\ref{fig:Focusmicromacrodyn}(c) shows raster plots of 200 randomly selected neurons. Without stimulation ($t<5$), most neurons spike coherently. Turning on the control at $t>5$ destroys the coherent spiking and stabilizes the initially unstable incoherent state.

\begin{figure}
\centering\includegraphics{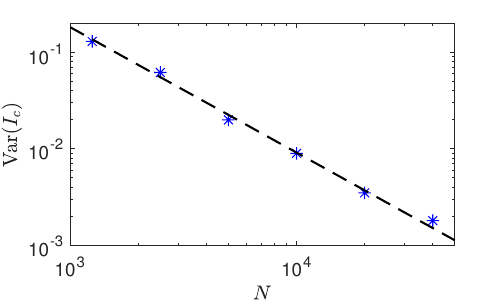}
\caption{The variance  $\mathrm{Var}(I_c)$ of the control perturbation in the post-transient regime as a function of the network size $N$. The asterisks show the result of the numerical simulation, and the dashed line shows the power-law approximation $\mathrm{Var}(I_c) =C N^{-\gamma}$ with $C=1450$ and $\gamma \approx 1.3$.}
\label{fig:VarvsN}
\end{figure}

Although the results of the mean-field equations and the microscopic model are very close, there is a fundamental difference in the asymptotic dynamics of these two models. As $t \to \infty$, the dynamic variables $(r,v)$ of the mean-field equations approach exactly the unstable fixed point $(r^*,v^*)$ of the uncontrolled system, and the control perturbation vanishes $I_c( t) \to 0$. In the microscopic model, the variables $(r,v)$ exhibit small fluctuations around the fixed point $(r^*,v^*)$, and the control perturbation $I_c(t)$ fluctuates around zero. Figure~\ref{fig:VarvsN} shows the dependence of the variance $\mathrm{Var}(I_c)$ of the control perturbation in the post-transient regime on the network size $N$. 
The variance decreases with increasing $N$ and vanishes at $N\to \infty$. This dependence is well described by the power law  $\mathrm{Var}(I_c) \sim N^{-\gamma}$ with $\gamma \approx 1.3$.    

\begin{figure}
\centering\includegraphics{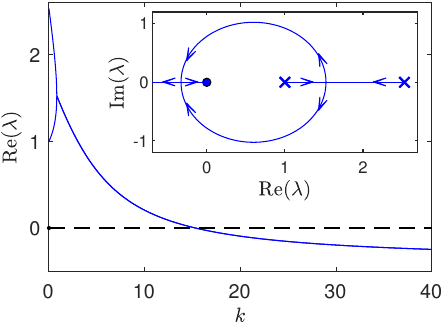}
\caption{Linear stability of a saddle incoherent state of a population of QIF neurons controlled by an unstable controller with a negative parameter $\omega_c=-1$. Dependence of two largest real parts of eigenvalues of the closed loop system on the control gain $k$. The inset shows the root loci of the characteristic Eq.~(\ref{Eigen}) in the complex plane $\lambda$ as $k$ changes from $0$ to $\infty$. The crosses on the real axes indicate the location of the eigenvalues at $k=0$, and the dot at the origin shows the location of one of the eigenvalues at $k = \infty$. Network parameters: $\Delta=1$, $V_{th}=50$, $\bar{\eta}=-7$, and $J=21$.}
\label{fig:Saddlelamk}
\end{figure}

Let us now consider the control of coherent oscillations near a homoclinic bifurcation. We will use the following set of the parameters: $\Delta=1$, $V_{th}=50$, $\bar{\eta}=-7$, and $J=21$. For these parameters, the free ($I_c=0$) system~(\ref{eqrv}) has a stable limit cycle and outside it a saddle point with coordinates $(r^*, v^*) \approx (0.4073,-0.3908)$ and two real eigenvalues $\lambda_{1,2} \approx (2.5306, -3.9255)$. Stabilization of the incoherent state associated with the saddle point cannot be attained with an ordinary LPF and requires the use of an unstable LPF with a negative parameter $\omega_c$. The eigenvalues of the saddle point in presence of the control are determined by the Eqs.~(\ref{Eigen}) and (\ref{Jacob}). The dependence of the two largest real parts of the eigenvalues on $k$ for a fixed $\omega_c=-1$ is shown in Fig.~\ref{fig:Saddlelamk}. 
The saddle point stabilization mechanism is best understood from the root loci diagram shown in the inset. Here we show the evolution of eigenvalues in the complex plane $\lambda$ as $k$ changes from $0$ to $\infty$. Two crosses on the real axes determine the location of the eigenvalues at $k=0$. One of them $\lambda=2.5306$ corresponds to a free network, and the other $\lambda=-\omega_c=1$ corresponds to a disabled unstable controller. With the increase of $k$, they approach each other on the real axes, collide and pass to the complex plane. At  $k\approx 15.3$, they cross symmetrically into the left half-plane (Hopf bifurcation). For very large $k\approx 91.8$, we have a collision on the real axis again, and then one of the roots goes to infinity, while the other approaches the origin. For $k >15.3$, the closed loop system is stable.

\begin{figure}
\centering\includegraphics{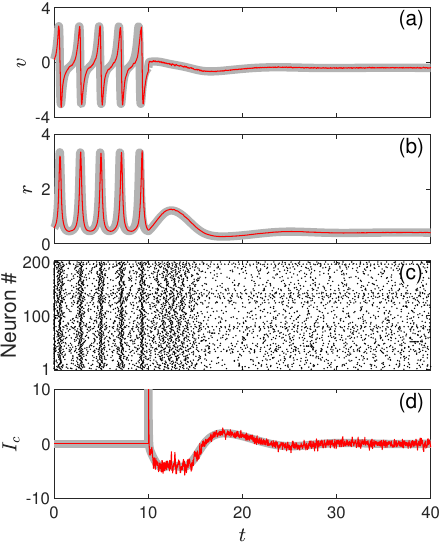}
\caption{Suppression of coherent oscillations  in a population of QIF neurons by stabilization of a saddle incoherent state with an unstable controller at $\omega_c=-1$ and $k=20$. 
As in Fig.~\ref{fig:Focusmicromacrodyn}, the dynamics derived from the mean-field equations are shown as thick gray curves, and the corresponding dynamics derived from the microscopic model of $10^4$ neurons are shown as thin red curves. All other designations are the same as in Fig.~\ref{fig:Focusmicromacrodyn}. The control turns on at $t=10$. The network parameters correspond to Fig.~\ref{fig:Saddlelamk}.}
\label{fig:Saddlemicromacrodyn}
\end{figure}

Figure~\ref{fig:Saddlemicromacrodyn} shows the results of stabilization of a saddle incoherent state with unstable controller parameters $\omega_c=-1$ and $k=20$.
As in Fig.~\ref{fig:Focusmicromacrodyn}, the dynamics derived from the mean-field equations are shown as thick gray curves, and the corresponding dynamics derived from the microscopic model of $10^4$ neurons are shown as thin red curves. Again, there is complete agreement between the mean-field theory and the microscopic theory. For $t<10$, there is no control, and the system is in the limit cycle mode, which is close to a homoclinic bifurcation. For $t>10$, the control is activated and the system approaches a stabilized incoherent saddle point. In the mean-field theory, the control perturbation vanishes asymptotically, while in the microscopic model it experiences small fluctuations around zero. Note that the steady-state spiking rate in saddle equilibrium is much lower than in focus equilibrium [cp. post-transient dynamics in Figs.~\ref{fig:Focusmicromacrodyn}(b) and \ref{fig:Saddlemicromacrodyn}(b)].

\section{Controlling two interacting populations
of excitatory and inhibitory QIF neurons} \label{sec:QIFtwo}
 
Let us now consider the control of a more complex network built from two connected populations of excitatory and inhibitory QIF neurons.  We follow the model discussed in Ref.~\cite{Pyragas2021} whose network architecture mimics the network architecture used in Parkinson's disease models. Such models are usually based on two interacting neural populations of the subthalamic nucleus (STN) consisting of excitatory neurons and the external segment of the globus pallidus  (GPe) consisting of inhibitory neurons (cf.,e.g., Ref.~\cite{Terman2002}). It was shown in ~\cite{Pyragas2021} that synchronous oscillations can be very effectively suppressed by HF stimulation of the inhibitory population, while HF stimulation of the excitatory population is ineffective. Here we want to test whether our control algorithm applied to the excitatory population can suppress synchronization.

The microscopic model of the network considered here is determined by the set of $2N$ neurons' membrane potentials  $\{V_j^{(E,I)}\}_{j=1,\ldots,N}$. They satisfy the system of $2N$ ODEs~\cite{Pyragas2021}:
\begin{eqnarray}
\tau_m\dot{V}_{j}^{(E,I)} &=&  ({V}_{j}^{(E,I)})^{2}+\eta_j^{(E,I)}+\mathcal{I}_{j}^{(E,I)},\nonumber\\
& & \;\; \text{if} \;\; {V}_{j}^{(E,I)}\ge V_{p} \;\; \text{then} \;\; {V}_{j}^{(E,I)} \leftarrow V_{r}, \label{modelEI}
\end{eqnarray}
where, $V_j^{(E,I)}$ is the membrane potential of neuron $j$ in the excitatory (E) or the inhibitory (I) population, and $\tau_m$ is the membrane time constant. The threshold potential assumption is the same as in the previous model: $V_{p}= -V_{r} = \infty$. 
The  heterogeneous parameters $\eta_j^{(E,I)}$ for populations E and I are taken  from two independent Lorentzian distributions:
\begin{equation}
g_{E,I}(\eta)=\frac{1}{\pi} \frac{\Delta_{E,I}}{(\eta-\bar{\eta}_{E,I})^2+\Delta_{E,I}^2}, \label{Lor}
\end{equation}
where $\Delta_{E,I}$ and $\bar{\eta}_{E,I}$ are respectively the width and the center of the distribution for the populations E and I. The last term $\mathcal{I}_{j}^{(E,I)}$ in Eqs.~(\ref{model}) describes synaptic coupling and external stimulation in the respective populations:
\begin{subequations}
\label{currents}
\begin{eqnarray}
\mathcal{I}_{j}^{(E)} & = & -J_{IE}r_I(t)+I_c(t),\label{currente}\\
\mathcal{I}_{j}^{(I)} & = & J_{EI}r_E(t)-J_{II}r_I(t).\label{currenti}
\end{eqnarray}
\end{subequations}
Unlike the previous model, here the interaction between neurons is provided by instantaneous pulses. Each time the potential of a given neuron reaches $\infty$, it resets to $-\infty$, and the neuron emits a Dirac delta spike, which contributes to the output of the network. The mean synaptic rates of E and I populations are as follows:
\begin{eqnarray}
r_{E,I}(t)=\lim_{\tau_s \to 0}  \frac{\tau_m}{\tau_s N} \sum_{i=1}^N \sum_k \int_{t-\tau_s}^t \delta(t'-(t_{i}^{k})_{E,I}) d t', \label{firingrate}
\end{eqnarray}
where $\delta(t)$ is the Dirac delta function and $(t_{i}^{k})_{E,I}$ is the time of the $k$th spike of the $i$th neuron in E and I population, respectively.  Parameters $J_{EI}$, $J_{IE}$ and $J_{II}$ denote synaptic weights. The current $J_{EI}r_E(t)$ excites I neurons due to the synaptic activity of E population and the current $-J_{IE}r_I(t)$ inhibits E neurons due to the synaptic activity of the I population. The current $-J_{II}r_I(t)$ recurrently inhibits neurons in population I. We are considering a stimulation protocol in which only the excitatory population is stimulated, so the control current $I_c(t)$ is only included in the Eq.~(\ref{currente}).

\begin{figure}
\centering\includegraphics{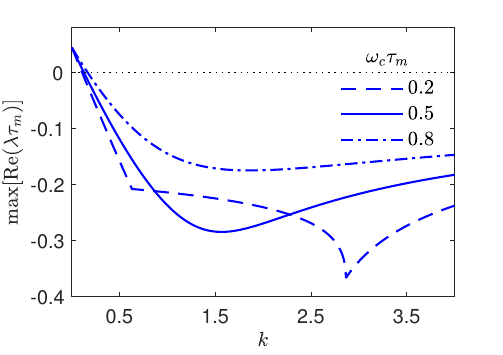}
\caption{Linear stability of incoherent state associated with a high-dimensional focus in a system of two interacting populations of excitatory and inhibitory QIF neurons in the presence of control. The entire network is controlled using the output and input of the excitatory population only. The maximum real part of the eigenvalues as a function of the control gain $k$ is shown for different values of the cutoff frequency $\omega_c$ of LPF. Network  parameters: $\Delta_E=0.05$, $\bar{\eta}_E=0.5$, $\Delta_I=0.5$, $\bar{\eta}_I=-4$, $J_{EI}=20$, $J_{IE}=5$, $J_{II}=0.5$, and $\tau_m=14$~ms.}
\label{fig:QIFSTNGPElamvsk}
\end{figure}

In the limit $N \to \infty$, this microscopic model reduces to an exact closed system of four ODEs for four biophysical quantities, mean firing rates $r_{E,I}$ and mean membrane potentials $v_{E,I}$ of populations E and I~\cite{Montbrio2015,Pyragas2021}:
\begin{subequations}
\label{eqrvEI}
\begin{eqnarray}
\tau_m\dot{r}_E & = & \Delta_E/\pi+ 2r_Ev_E, \label{eqrE}\\
\tau_m\dot{v}_E & = & \bar{\eta}_E +v_E^2-\pi^2 r_E^2-J_{IE}r_I+I_c(t),\label{eqvE}\\
\tau_m\dot{r}_I & = & \Delta_I/\pi+ 2r_Iv_I, \label{eqrI}\\
\tau_m\dot{v}_I & = & \bar{\eta}_I +v_I^2-\pi^2 r_I^2+J_{EI}r_E-J_{II}r_I. \label{eqvI}
\end{eqnarray}
\end{subequations}
Bifurcation analysis of an uncontrolled ($I_c=0$) system~(\ref{eqrvEI}) showed a wide variety of different dynamic modes~\cite{Pyragas2021}. Here we focus on the case when the system has a single attractor, the limit cycle. Specifically, we consider the following set of system parameters: $\Delta_E=0.05$, $\bar{\eta}_E=0.5$, $\Delta_I=0.5$, $\bar{\eta}_I=-4$, $J_{EI}=20$, $J_{IE}=5$, $J_{II}=0.5$, and $\tau_m=14$~ms. At these parameters, the system, along with a stable limit cycle, has an unstable fixed point, which is a high dimensional focus with coordinates\\
$
(r_E^*,v_E^*,r_I^*,v_I^*) \approx (0.1319, -0.0603, 0.0663, -1                   .1990)
$\\
and two pairs of complex conjugate eigenvalues
 $\lambda_{1,2} \approx (0.0448 \pm 1.0304 i)/\tau_m$ and  $\lambda_{3,4} \approx (-2.5634 \pm 0.8190 i)/\tau_m$.
 Our goal is to stabilize this fixed point using the control algorithm defined by Eqs.~(\ref{controllaw}), with the constraint that the available network output is the mean membrane potential of the excitatory population, $v=v_E$, and the control current $I_c$ is applied only to the excitatory population.  Linear stability of the fixed point in the presence of control can be analyzed in a similar way as in the previous model. Now the characteristic equation has five eigenvalues. The dependence of the $\max[\Re(\lambda)]$ on the control gain $k$ for three different values of the cutoff frequency $\omega_c$ is shown in Fig.~\ref{fig:QIFSTNGPElamvsk}. 
Again we see that the stability condition  $\max[\Re(\lambda)]<0$ is satisfied in a wide range of the control parameters $k$ and $\omega_c$.

\begin{figure}
\centering\includegraphics{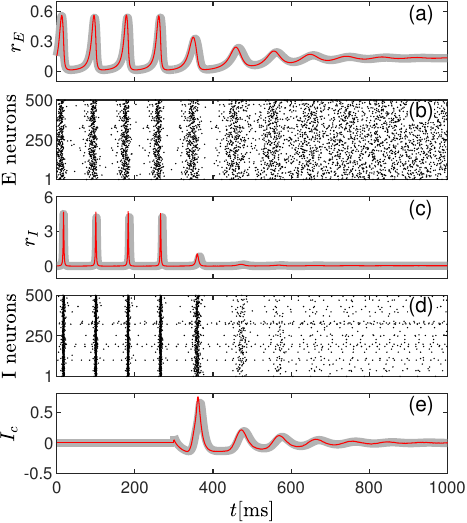}
\caption{Suppression of coherent oscillations in a system of two interacting populations of excitatory and inhibitory QIF neurons by stabilization of unstable incoherent state associated with a high-dimensional focus. Dynamics of mean spiking rate of (a)  excitatory and (c) inhibitory populations, and (e) control perturbation applied to the excitatory population. The dynamics derived from the mean-field equations are shown as thick gray curves, and the corresponding dynamics derived from the microscopic model with $10^4$ neurons in each excitatory and inhibitory population are shown as thin red curves.  (b), (d) Raster plots of 500 randomly selected neurons in E and I populations, respectively. The control turns on at $t=300$ ms. Network parameters as in Fig.~\ref{fig:QIFSTNGPElamvsk}. Controller parameters: $\omega_c =0.5/\tau_m$ and $k=0.5$.}
\label{fig:ControlDynEI}
\end{figure}

Figure~\ref{fig:ControlDynEI} shows the performance of the control algorithm for fixed values of $\omega_c=0.5/\tau_m$ and $k=0.5$.
The dynamics of the free ($t< 300$ ms) and controlled ($t> 300$ ms) network, obtained from the mean-field Eqs.~(\ref{eqrvEI}), are shown as thick gray curves.   The control switches the state of the system from coherent limit cycle oscillations to the stabilized incoherent state and the feedback perturbation asymptotically vanishes. These results are consistent with numerical simulations of a microscopic model with $N=10^4$ neurons in each excitatory and inhibitory population (thin red curves). As in the previous case, we changed the variables
\begin{equation}
V_j^{(E,I)} = \tan(\theta_j^{(E,I)}/2)\label{eqtransftet1}
\end{equation}
to rewrite the Eqs.~(\ref{modelEI})  in terms of theta neurons: 
%
\begin{eqnarray}
\tau_m \dot{\theta}_{j}^{(E,I)}&=& 1-\cos \left(\theta_{j}^{(E,I)}\right)\nonumber\\
&+&\left[1+\cos \left(\theta_{j}^{(E,I)}\right)\right]\left[\eta_{j}^{(E,I)}+\mathcal{I}_{j}^{(E,I)} \right]. \label{thetaEI}
\end{eqnarray}
%
We integrated these equations by the Euler method with a time step of $d t = 2\times 10^{-5}$.  For the numerical implementation of Eq.~(\ref{firingrate}), we set $\tau_s =  5\times 10^{-5} \tau_m$.  To estimate the variables of the mean-field theory, we calculated the Kuramoto order parameters
\begin{equation}
\label{eqZEI}
Z_{E,I}=\frac{1}{N}\sum\limits_{j=1}^{N}\exp(i \theta_j^{(E,I)})
\end{equation} 
for each population and evaluated the mean spiking rates and mean membrane potentials for populations E and I as~\cite{Montbrio2015}: 
\begin{equation}
\label{eqWEI}
r_{E,I}=\frac{1}{\pi}\operatorname{Re}\left(\frac{1-Z_{E,I}^*}{1+Z_{E,I}^*}\right),  \ v_{E,I}=\operatorname{Im}\left(\frac{1-Z_{E,I}^*}{1+Z_{E,I}^*}\right),
\end{equation} 
where $Z_{E,I}^*$ denotes complex conjugate of $Z_{E,I}$. Panels (a), (c) and (e) in Fig.~\ref{fig:ControlDynEI} show a good agreement of time traces obtained from mean filed equations and microscopic model. Panels (b) and (d) are raster plots of randomly selected $500$ neurons in excitatory and inhibitory populations, respectively.

\section{Controlling a population of chaotically spiking Hindmarsh-Rose neurons}\label{sec:HindMarshRose}

As a final example, consider the control of synchronous oscillations in a heterogeneous population of electrically coupled Hindmarsh-Rose neurons~\cite{Hindmarsh1984}:
\begin{subequations}
\label{eqHR}
\begin{eqnarray}
\dot{v}_j & = & y_j -v_j^3+3v_j^2-z_j+I_j+K(v-v_j)+I_c(t),\label{eqHRv}\\
\dot{y}_j & = &  1-5v_j^2-y_j,\label{eqHRy}\\
\dot{z}_j & = & r[\nu (v_j-\kappa)-z_j], \quad j=1,\ldots, N.\label{eqHRz}
\end{eqnarray}
\end{subequations}
Here, $v_j$, $y_j$ and  $z_j$ are the membrane potential, the spiking variable and the adaptation current of the $j$th neuron, respectively. The variable
\begin{equation}
v = \frac{1}{N}\sum_{i=1}^N v_i\label{eqHRvmean}
\end{equation}
is the mean membrane potential. The heterogeneity of neurons is provided by currents $I_j$, which we randomly select from a Gaussian distribution with a mean value of $3$ and a variance of $0.1$. Parameters $r=0.06$, $\nu=4$ and $\kappa=-1.56$ are chosen such that free ($K=0$ and $I_c=0$) neurons generate chaotic bursts. The term $K(v-v_j)$ in the Eq.~(\ref{eqHRv}) determines the electrical coupling between neurons, where $K$ is the coupling strength. To get synchronized oscillations of the uncontrolled population, we take this parameter large enough, $K=0.1$. The last term $I_c(t)$ in this equation is the control current given by Eqs.~(\ref{controllaw}). 

\begin{figure*}
\centering\includegraphics{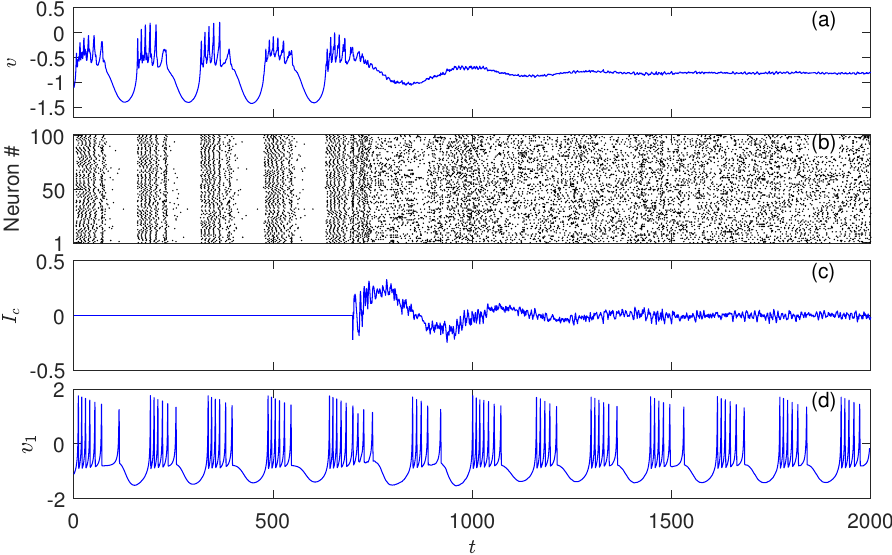}
\caption{Suppression of coherent oscillations in a population of electrically coupled Hidmarsh-Rose neurons. The control is activated at the time $t=700$. (a) Dynamics of the mean membrane potential. (b) Raster plot of $100$ randomly selected neurons. (c) and (d) Time traces of the control perturbation and the membrane potential of the first neuron, respectively. Network parameters: $r=0.06$, $\nu=4$, $\kappa=-1.56$, $K=0.1$ and $N=10^4$. Heterogeneous currents $I_j$ in Eq.~(\ref{eqHRv}) are randomly selected from a Gaussian distribution with a mean value of $3$ and a variance of $0.1$. Controller parameters: $\omega_c=0.05$ and $k=2$.}
\label{fig:chaosmicro}
\end{figure*}

Figure~\ref{fig:chaosmicro} shows how the control with fixed parameters $\omega_c=0.05$ and $k=2$ changes the dynamics of a population of $N=10^4$ coupled neurons. Without control ($t <700 $), synchronous oscillations of large amplitude are observed in the dynamics of the mean membrane potential $v(t)$, and coherent bursts are visible on the raster plot. Activation of control at $t>700$ effectively suppresses synchronous oscillations of the mean membrane potential, and neurons demonstrate incoherent bursts. As in previous examples, only small amplitude oscillations around zero are observed in the asymptotic dynamics of control perturbation $I_c(t)$. Figure~\ref{fig:chaosmicro}(d) demonstrates that control almost does not affect the amplitude dynamics of individual neurons. As an example, we show the time trace of the membrane potential of the first neuron $v_1(t)$ before and after activation of control. 

Note that, unlike the previous examples, there is no known way to reduce this model to a low-dimensional system. Thus, here we cannot theoretically estimate the mean value of the membrane potential of an unstable incoherent state in the thermodynamic limit and determine whether the equilibrium  is associated with an unstable focus or  saddle and how its stability changes in the presence of control.
However, our algorithm does not require such detailed knowledge and, with an appropriate choice of control parameters, works just as well as in previous relatively simple models that allow a low-dimensional reduction in the thermodynamic limit. Numerical simulations of this model show that our algorithm works only when $\omega_c>0$ and fails when $\omega_c<0$. This allows us to conclude that the unstable equilibrium in this model is an unstable focus.

\section{\label{sec:Conclusions} Conclusions}
 
We considered the problem of suppressing collective synchronous oscillations in large-scale neural networks. This problem is relevant in neurology, as excessive synchronized oscillations in certain areas of the brain are often associated with various neurological disorders~\cite{Hammond2007,Jiruska2013,Gerster2020,Tass2012tin}. Synchronized oscillations usually appear when an equilibrium incoherent state of the network becomes unstable. Information about unstable network states is difficult to extract from experimental data. We have shown that a priory unknown unstable incoherent states of large-scale neural networks can be effectively stabilized using a simple first order feedback controller based on a low-pass filter. Initially, this controller was developed for  stabilization of unknown unstable equilibrium points of low-dimensional dynamical systems~\cite{Pyragas2002,Pyragas2004} and has not yet been tested for high-dimensional systems such as neural networks, consisting of a huge number of interacting neurons. 

We have demonstrated the effectiveness of our control algorithm on three examples of neural networks. The first two examples refer to QIF neurons. In the thermodynamic limit, microscopic models of networks built from QIF neurons can be reduced to exact low-dimensional systems of mean-field equations. This greatly simplifies the analysis of the effect of control on network dynamics. In the first example, we demonstrated the suppression of synchronous oscillations in a population of excitatory QIF neurons interacting through synaptic pulses of finite width. Here we have stabilized two types of incoherent states associated with an unstable focus and a saddle equilibrium point. Until now, the control of the saddle equilibrium state in neural networks has not been considered in the literature. Here we have achieved stabilization of the saddle state with the help of an unstable controller. In the second example, we considered the control of a network built from two connected populations of excitatory and inhibitory QIF neurons, whose architecture mimics that, used in Parkinson's disease models~\cite{Terman2002}. Previously, it was shown that high-frequency stimulation of the inhibitory population can effectively suppress synchronization in such a network, but stimulation of the excitatory population is ineffective~\cite{Pyragas2021}. Here, our algorithm provided effective stabilization of the incoherent state of the network by using the output and input of the excitatory population. For both first examples, the results derived from the mean-field equations were confirmed by numerical simulations of the respective microscopic models. We have shown that networks of $10^4$ neurons are quantitatively well described by mean-field equations. In the third example, we demonstrated the suppression of coherent oscillations in a population of electrically coupled Hindmarsh-Rose neurons. Low-dimensional reduction of the equations of the microscopic model is impossible in this case. However, the direct application of the control algorithm to the microscopic model showed that it works just as well as in the previous two examples. Note that successful stabilization of unstable incoherent states makes them experimentally observable, and this can serve as a quantitative benchmark for assessing the quality of neural network models.

Finally, we summarize the main advantages of the proposed algorithm for suppressing coherent oscillations in large-scale neural networks: (i) the algorithm does not require any detailed knowledge of the network model and its unstable incoherent equilibria; (ii) the algorithm is robust to changes in control parameters; (iii) the algorithm can stabilize not only incoherent states associated with an unstable focus but also with a saddle equilibrium point; (iv) for large networks the algorithm is weakly invasive: the control perturbation decreases according to a power law with increasing network size and vanishes as the network size tends to infinity; (v) the algorithm is adaptive, which means that it provides tracking of the equilibrium states in the case of slowly varying system parameters (see~\cite{Pyragas2002,Pyragas2004} for details).  

In this paper, we limited ourselves to the consideration of the simplest first-order controller to stabilize unknown incoherent states. More complex networks may require higher-order generalized adaptive controllers~\cite{Pyragas2002}. In addition, we emphasize that mean-field equations derived from microscopic dynamics accurately describe synchronization processes in large networks, and these models are well suited for testing and developing various algorithms for suppressing unwanted coherent oscillations.

\section*{Acknowledgments}

This work is supported by grant No. S-MIP-21-2 of the Research Council of Lithuania. 

\bibliographystyle{elsarticle-num}
\bibliography{adaptive}
\end{document}